\documentstyle[epsfig]{article}
\def\be{\begin{equation}}
\def\ee{\end{equation}}
\def\ba{\begin{eqnarray}}
\def\ea{\end{eqnarray}}
\def\de{\partial}
\def\Br{\langle}
\def\kt{\rangle}
\newcommand{\eq}{\begin{equation}}
\newcommand{\en}{\end{equation}}
\newcommand{\eqa}{\begin{eqnarray}}
\newcommand{\ena}{\end{eqnarray}}
\newcommand{\bea}{\begin{eqnarray}}
\newcommand{\eea}{\end{eqnarray}}

\newcommand{\um}{\frac12}

\newcommand{\ZZ}{\hbox{{\rm Z{\hbox to 3pt{\hss\rm Z}}}}}

\newcommand{\NP}[1]{Nucl.\ Phys.\ {\bf #1}}
\newcommand{\PL}[1]{Phys.\ Lett.\ {\bf #1}}

\newcommand{\PR}[1]{Phys.\ Rev.\ {\bf #1}}
\newcommand{\PRL}[1]{Phys.\ Rev.\ Lett.\ {\bf #1}}

\newcommand{\IJMP}[1]{Int.\ J.\ Mod.\ Phys.\ {\bf #1}}

\begin{document}
\begin{titlepage}
\begin{flushright}
DFTT 45/96\\
HUB-EP-96/51\\
CS-TH 4/96\\
September 1996
\end{flushright}
\vskip0.5cm
\begin{center}
{\Large\bf  String Effects in the Wilson Loop:}
\vskip0.2cm
{\Large\bf   a high precision numerical test}
\end{center}
\vskip 0.6cm
\centerline{M. Caselle$^a$, R. Fiore$^b$, F. Gliozzi$^a$,}
\centerline{M. Hasenbusch$^c$ and P.Provero$^a$}
\vskip 0.6cm
\centerline{\sl  $^a$ Dipartimento di Fisica
Teorica dell'Universit\`a di Torino}
\centerline{\sl Istituto Nazionale di Fisica Nucleare, Sezione di Torino}
\centerline{\sl via P.Giuria 1, I--10125 Torino, Italy
\footnote{e--mail: caselle, gliozzi, provero~@to.infn.it}}
\vskip .2 cm
\centerline{\sl $^b$ Dipartimento di Fisica, Universit\`a della Calabria}
\centerline{\sl Istituto Nazionale di Fisica Nucleare, Gruppo collegato
di Cosenza}
\centerline{\sl Rende, I--87030 Cosenza, Italy
\footnote{e--mail: fiore~@fis.unical.it}}
\vskip .2 cm
\centerline{\sl  $^c$ Humboldt Univerit\"at zu Berlin, Institut f\"ur 
Physik}
\centerline{\sl Invalidenstr. 110, D--10099 Berlin, Germany 
\footnote{e-mail: hasenbus@birke.physik.hu-berlin.de}}
\vskip .2 cm
\vskip 0.6cm
\begin{abstract}
We test numerically the effective string 
description of the infrared limit of lattice gauge theories in the 
confining regime. We consider the $3d$ $\ZZ_{2}$ lattice gauge theory, 
and we define ratios of Wilson loops such that the predictions of the 
effective string theory do not contain any adjustable parameters. In 
this way we are able to obtain a degree of accuracy high enough to 
show unambiguously that the flux--tube fluctuations are described, in 
the infrared limit, by an effective 
 bosonic string theory.
\end{abstract}
\end{titlepage}
\setcounter{footnote}{0}
\def\thefootnote{\arabic{footnote}}
\section{Introduction}

It is widely believed that the confining regime of Yang-Mills theory is
described by some kind of effective string model~\cite{conj}.
 It is the aim of this paper to present some numerical evidence
  supporting this conjecture and, what is most
important,
to study the nature of the effective string which is actually realized
in the infrared regime of Lattice Gauge Theories (LGTs). 

This conjecture has by now a very long history. It originates
from two independent observations. The first one is of 
phenomenological nature,
and dates  before the formulation of QCD. 
It is related to the observation that the linearly rising Regge
trajectories in meson spectroscopy can be easily explained assuming a
string-type interaction between the quark and the antiquark. This observation
was at the origin 
of a large amount of papers which
tried to give a consistent quantum description of strings. 

In the same years, a completely independent 
observation shed some new light on
the problem of the confinement of quarks. It was shown that in the strong
coupling limit of pure LGTs the
interquark potential rises linearly, 
and that the chromoelectric flux lines are
confined in a thin ``string like'' flux tube~\cite{wilson}; however this
approximation is plagued by lattice artifacts which make it inadequate for the
continuum theory.
Some clear indications were later found that the vacuum expectation value of
Wilson loops could be rewritten as a string functional integral even in the
continuum \cite{polya,gene,nambu}. This led to conjecture  that there exists 
an exact duality between gauge fields and strings \cite{polya}.
Although this approach has been studied extensively 
(see e. g. \cite{mm,awa,pol,gross}), 
such an exact duality has been proven only in 
$2d$ QCD in the large $N$ limit. In this paper we do not assume this strong 
conjecture, but a very mild version of it, which states that the
behavior of large Wilson loops is described in the infrared limit by an
effective two-dimensional field theory which accounts for the string-like
properties of long chromoelectric flux tubes. 
We shall refer to such a $2d$ field
theory in the following as the ``effective string theory''.

Let us  briefly outline the pattern of this paper.
In the next section we shall discuss some general  
features of the effective string theory.  In sect. 3
we shall evaluate in this framework
the finite size correction for the expectation values of Wilson loops.
This theoretical prediction will then be compared with the result of
a set of high precision Monte Carlo simulations in sect.5, while
in sect.4 we shall
describe the $3d$ $\ZZ_2$ gauge model and the 
algorithms that we used to measure these expectation
values: a gauge version of the microcanonical demon
algorithm and a non-local cluster algorithm applied to the dual spin model.
Finally sect. 6 will be devoted to some concluding remarks.

\section{ Some features of the effective string theory}
The aim of this section is to give some
introductory material so as to make this paper self--contained. We shall first
show how the effective string emerges in LGT (sect. 2.1) and discuss the
peculiar finite size effects 
that it induces in the expectation value of large Wilson loops 
(sect. 2.2). We shall then address 
the question of the ``string universality''
namely the fact that these effects show a substantial independence on the
gauge group (sect. 2.3). This will allow us to concentrate on the simplest
possible non trivial gauge theory, namely the $3d$ gauge 
Ising model to test the
effective string theory. By means of a duality transformation we shall 
then use some
results from the physics of interfaces to obtain some
preliminary expectation on the nature of the effective string model 
(sect. 2.4).
These expectations will then be confirmed in the rest of the paper by a direct
comparison with the results of Monte Carlo simulations.

\subsection{ Flux Tubes in the Rough Phase}

The confining regime of LGTs contains in general two phases: the 
strong coupling phase and the rough
phase. The two are separated by the so called ``roughening transition'' 
which is the point in which the  strong coupling expansion of the Wilson loop
ceases to converge~\cite{rough,lsw}.
 These two phases are related to
two different behaviors of the quantum fluctuations of the 
flux tube around its
equilibrium position~\cite{lsw}. 
In the strong coupling phase, these fluctuations are
massive, while in the rough phase they become massless 
and hence survive in the
continuum limit. This fact has several
consequences:
\begin{description}
\item{(a)} The flux--tube fluctuations can be described by a suitable
two-dimensional massless quantum field theory (QFT), 
where the fields describe the
transverse displacements of the flux tube. This quantum field theory is 
expected to be very
complicated and will contain in general non renormalizable interaction terms
\cite{lsw,cfghpv}. 
However, exactly because these interactions are non-renormalizable, their
contribution becomes negligible in the infrared limit (namely for large Wilson
loops). In this infrared limit the QFT becomes a conformal invariant 
field theory (CFT).
\item{(b)} The massless quantum fluctuations delocalize the flux tube
 which acquires a nonzero width, which diverges logarithmically as the
 interquark distance increases~\cite{lmw, width}.
\item{(c)} The quantum fluctuations
 give a non-zero contribution to the interquark potential,
which is related to the partition function of the above $2d$ QFT.
Hence if the $2d$ QFT is simple enough to be exactly solvable (and this is in
general the case for the CFT in the infrared limit) also these
contributions can be evaluated exactly.
\item{(d)} In the simplest case, this CFT  is simply
the two dimensional conformal field theory 
of $(d-2)$ free bosons ($d$ being the number of spacetime dimensions 
of the original gauge model); its exact solution will be discussed below.

\end{description}

\vskip 0.5cm
\subsection{ Finite Size Effects}
\vskip 0.1cm
The feature of the effective string description which is best
 suited to be studied by numerical methods is the presence of 
finite--size effects. 
Wilson loops are classically expected to obey the famous 
area-perimeter-constant law. 
This means that the expectation value of a Wilson
loop of size $R\times T$ must behave, as $R$ and $T$ vary, as:

\be
<W(R,T)>=e^{-\sigma RT+p(R+T)+k}~.
\label{area}
\ee

This law is indeed very well 
verified in the strong coupling regime (before the
roughening transition). However in the rough phase it must be modified.
One must multiply it by the 
partition function of the $2d$ QFT describing  the quantum
fluctuations of the flux tube. As we have seen before, this QFT 
in the infrared limit becomes a two dimensional 
CFT, and its partition function $Z_{q}(R,T)$
 can be evaluated exactly.
(See {\it e.g.} Ref.~\cite{cft}
 for a comprehensive review on  CFTs). In the next section we shall
describe in detail an example of this type of calculations.
Eq.~(\ref{area}) in the rough phase becomes:

\be
<W(R,T)>=e^{-\sigma RT+p(R+T)+k}Z_{q}(R,T)~.
\label{quantum}
\ee

In the limit $T>>R$ (which is the relevant one to extract the interquark
potential) the partition function of the most general CFT yields:

\be
\lim_{T\to\infty} \frac{1}{T}\log Z_{q}(R,T)= \frac{c\pi}{24 R}~~,
\label{limit}
\ee
where $c$ is the central charge of the CFT. In the simplest possible case,
namely when the CFT describes a collection of $n$ free bosonic fields, 
we have
$c=n$. Thus for the free boson realization of the effective string 
theory, 
we find $c=d-2$. This is the result obtained by
L\"uscher, Symanzik and Weisz in~\cite{lsw}.

The interquark potential is thus given by:
\be
V(R)=-\lim_{T\to\infty} \frac{1}{T}\log\Br W(R,T)\kt
=\sigma R - \frac{c\pi}{24 R}~~.
\label{luescher}
\ee

The $1/R$ term in the potential is the finite size effect mentioned above; 
it is
completely due to the quantum fluctuations of the flux tube and, if
unambiguously detected, it represents a strong evidence 
(the strongest we have) 
in favor of the effective string picture discussed above. Moreover if the 
measurement
is precise enough we can in principle extract numerically the value of $c$ and
thus select which kind of effective string model describes the infrared 
regime of the 
LGT under examination. 

In LGTs with continuous gauge groups the interquark potential
 has  a further term which is due  to the one gluon exchange. It can be
 evaluated perturbatively, and it
 exists only in the ultraviolet regime, namely for
 small Wilson loops. It has the form of a Coulombic 
interaction, hence in (3+1)
 dimensions it is also of the type 
$1/R$ and can shadow the contribution coming
 from the flux tube fluctuations. 
 Even if it holds only in the perturbative regime, we cannot
fix a sharp threshold after which it disappears, so it could well be that,
 in the
set (of large Wilson loops) from which we extract our data
we find
a superposition of the two terms. There are two ways
to avoid this problem:
\begin{description}
 \item{(a)}
  Study LGT in three dimensions where the perturbative term has 
a $\log R$ form
  instead of $1/R$, and does not mix up with the string contribution.
 \item{(b)}
Study Wilson loops with comparable values of $T$ and $R$. In this case, 
  the whole
functional form of the two interaction terms becomes important. These are
completely different and thus can be separated.
\end{description}

Actually the latter option 
is anyway much more interesting than the study of the
$T\to\infty$ limit only, since the whole functional form of the CFT partition
function is richer, and allows more stringent 
numerical tests. In this 
paper  we shall follow this strategy and study always
Wilson loops with comparable values of $T$ and $R$. This choice will magnify
the finite size effects and will simplify our numerical work. 

Since the beginning of eighties several numerical works have been done to study
this problem. The main results can be summarized as follows:

 \begin{description}
  \item{(a)} A $1/R$ term 
exists in the potential. In the case of (3+1) LGT with
  continuous gauge group it can be separated form the perturbative term.
  \item{(b)} The central charge has not been measured with good
  precision, however the numbers seem to be in reasonable agreement with the
  $c=d-2$ prediction of L\"uscher Symanzik and Weisz. 
The lack of precision, and
  the consequent impossibility to determine which kind of effective string is
  actually realized in the infrared regime of LGT was one of the main 
motivations of the present work. 
  \item{(c)} The same correction is found in very different LGTs, ranging from
  the $3d$ Ising gauge model to the $4d$ SU(3) model. 
This remarkable universality
  is an important feature of these finite size effects of the effective string
  description.

\end{description}

\subsection{ String Universality}

As a matter of fact not only the string corrections, but also other 
features of the  infrared regime of
LGTs  in the confining phase  display a high
degree of universality, namely they seem not to
 depend on the choice of the gauge group. This is the case for instance of the
 ratio between the critical temperature and the square root of the string
 tension, or the behavior of the spatial string tension above the deconfinement
 transition.  
Recently this same universality has been evidentiated in 
the pattern of the glueball states for various three dimensional gauge 
models \cite{glue}.
All these examples show a substantial independence on the gauge group and a
small and smooth dependence on the number of spacetime dimensions.

This ``experimental fact''
 has a natural explanation in the context of an effective string model:
even if in principle different gauge models could be described
 by different string theories\footnote{Notice however that recently
 Polyakov~\cite{pol} has
 put forward the conjecture that, even at the fundamental level all the Yang
 Mills models are described by one and the same string theory.},
 in the infrared  regime, as the interquark distance increases 
all these different string theories 
flow toward the common fixed point which is not anomalous and corresponds, 
in the simplest case discussed above,
to the two dimensional conformal field theory 
of $(d-2)$ free bosons.
Also the small dependence on the number of spacetime 
dimensions of the theory is
well predicted by the effective string theory.
 
 A very important consequence of this universality is that it allows us
to study the universal infrared behavior of Wilson loops 
  in the case of the three dimensional 
Ising or $\ZZ_2$ gauge model, which is the simplest 
non-trivial LGT and allows high
precision Monte Carlo simulations with a relatively small amount of CPU time.
We already used this opportunity in some previous works on the subject, and
shall again use it in the present paper. The simplicity of the model and the
construction of a new very powerful algorithm to simulate it will allow us to
reach a very high degree of precision, 
never reached before even in the context
of the Ising model.
This will allow us to test the predictions of the effective string 
theory with 
unprecedented confidence and to determine unambiguously the
 nature of the effective string which describes the infrared regime of LGT.

\vskip 0.5cm
\subsection{ Interfaces}
\vskip 0.1cm
The physics of the Wilson loop is closely related to the behavior of 
interfaces in spin models, which is known to be described by an 
effective string theory, 
the only difference being in the boundary conditions.

Indeed the most precise numerical test
 of the  effective string theory up to now has been performed  in the physics 
 of interfaces 
of the three dimensional Ising model \cite{cfghpv}. 
The gauge Ising model is transformed by 
duality in the spin Ising model and the Wilson loops  are related to interfaces 
in the context of the spin model\footnote{The term 
``roughening transition'' was actually first introduced in 
the context of the physics of interfaces and only later it was also used in the
study of Wilson loops.}. The string tension is dual to the interface
tension.
 The effective string model is known, in the 
context of interface physics as the capillary wave model.
According to this model interfaces in the rough phase are described by
a $2d$ quantum field theory of a free, massless bosonic 
field plus an infinite set
of non-renormalizable interaction terms; these can be written explicitly in
terms of the bosonic field and of its derivatives. In the infrared limit the
capillary wave model becomes a bosonic free field theory hence a CFT with $c=1$.
In a set of recent papers  this prediction has been compared with some
Monte Carlo result on the interface free energy in the $3d$ Ising model and a
remarkable agreement has been found.
\\
Moreover, this same effective description of interfaces has been 
derived analitically in the $3d$ $\lambda\phi^{4}$ quantum field theory
\cite{pv2}. The high degree of universality of this picture is shown 
by the fact that also interfaces in the $3d$ three-state Potts model 
are accurately described by the same effective theory  \cite{pv1}.
\\
It is therefore quite natural to conjecture that Wilson loops in the 
infrared limit are described by the same CFT.
The aim of this work is precisely to verify whether the effective 
string description 
that has been shown to describe interface physics in $3d$ spin 
models can be extended to the Wilson loop behavior in the 
corresponding gauge model, once the necessary modifications in the 
boundary conditions have been taken into account.
It turns out that 
even in this case the quantum fluctuations in the infrared limit are
described by a $c=1$ CFT.

\section{The prediction of the effective string theory}
In this section we review the computation of the effective string theory
prediction for the behavior of the Wilson loop in the infrared limit: 
we will show how a rather large class of bosonic string theories 
reduce in this limit to the theory of 
$d-2$ free massless scalar fields, and we will compute the partition 
function $Z_{q}(R,T)$ of this CFT. This expression will be plugged in 
Eq.~(\ref{quantum}) to obtain the prediction of the effective string 
theory for the behavior of large Wilson loops \cite{ambj}.

Let us consider for example the Nambu string action,
given by the area of the 
world--sheet:
\be
S=\sigma\int_0^{T}d\tau\int_0^{R} d\varsigma\sqrt{g}\ \ ,\label{action}
\ee
where $g$ is the determinant of the two--dimensional metric induced on
the world--sheet by the embedding in $R^d$:
\ba
g=\det(g_{\alpha\beta})&=&\det\ \de_\alpha X^\mu\de_\beta X^\mu\ \ .\\
&&(\alpha,\beta=\tau,\varsigma,\ \mu=1,\dots,d)\nonumber
\ea
 and $\sigma$ is the string tension.
\par
The reparametrization and Weyl invariances of the action (\ref{action})
require a gauge choice for quantization. We choose the "physical gauge"
\ba
X^1&=&\tau\nonumber\\
X^2&=&\varsigma
\ea
so that $g$ is expressed as a function of the transverse degrees of freedom
only:
\ba
g&=&1+\de_\tau X^i\de_\tau X^i+\de_\varsigma X^i\de_\varsigma X^i\nonumber\\
&&\ \ \ +\de_\tau X^i\de_\tau X^i\de_\varsigma X^j\de_\varsigma X^j
-(\de_\tau X^i\de_\varsigma X^i)^2\\
&&\ \ \ \ (i=3,\dots,d)~.\nonumber
\ea
The fields $X^i(\tau,\varsigma)$ satisfy Dirichlet boundary conditions on $M$:
\be
X^i(0,\varsigma)=X^i(T,\varsigma)=X^i(\tau,0)=X^i(\tau,R)=0\ \ .
\ee
Due to the Weyl anomaly this gauge choice can be performed at the 
quantum level only in the
critical dimension $d=26$. However, the effect of
the anomaly is known to disappear at large distances \cite{olesen}, 
which is the region we are interested
in.\par
Expanding the square root in Eq.~(\ref{action}) we obtain, discarding 
terms of order $X^4$ and higher
\ba
S&=&\sigma R T+\frac{\sigma}{2}\int d^2\xi X^i (-\de^2) X^i\label{free}\\
\de^2&=&\de_\tau^2+\de_\varsigma^2~.
\ea
It is easy to see that this expansion of the action corresponds, for 
the partition function, to an expansion in powers of $(\sigma R T)^{-1}$.
Therefore the action (\ref{free}) describes the infrared limit of the 
model defined by Eq.~(\ref{action}), and will be relevant to the 
physics of large Wilson loops. While 
the choice of the action (\ref{action}) is not universal, a large 
class of bosonic effective string models reduce to this CFT in the infrared 
limit.The contribution of the fluctuations of the flux--tube to the Wilson 
loop expectation value in the infrared limit will be the partition 
function of our CFT, given by
\be
Z_{q}(R,T)\propto 
\left[\det(-\de^{2})\right]^{-\frac{d-2}{2}}~.\label{z1loop}
\ee
\vskip0.5cm\noindent
The determinant must be evaluated with Dirichlet boundary conditions. 
\par
The spectrum of $-\de^2$ with Dirichlet boundary conditions is 
given by the eigenvalues
\be
\lambda_{mn}=\pi^2\left(\frac{m^2}{T^2}+\frac{n^2}{R^2}\right)
\ee
corresponding to the normalized eigenfunctions
\be
\psi_{mn}(\xi)=\frac{2}{\sqrt{R T}}\sin\frac{m\pi\tau}{T}
\sin\frac{n\pi\varsigma}{R}~.
\ee
The determinant appearing in Eq.~(\ref{z1loop}) can be regularized with the 
$\zeta$-function technique: defining
\be
\zeta_{-\de^2}(s)\equiv\sum_{mn=1}^\infty\lambda_{mn}^{-s} \ \,\label{zeta}
\ee
the regularized determinant is defined through the analytic continuation
of $\zeta_{-\de^2}^\prime(s)$ to $s=0$:
\be
\det(-\de^2)=\exp\left[-\zeta_{-\de^2}^\prime(0)\right]\ \ .
\ee
\par
The series in Eq.~(\ref{zeta}) can be transformed, using the Poisson summation
formula, to read
\ba
&&\zeta_{-\de^2}(s)=-\frac{1}{2}\left(\frac{R^2}{\pi^2}\right)^s
\zeta_R(2s)+\frac{\sqrt{\pi}Im\tau\Gamma(s-1/2)}
{2\Gamma(s)}\left(\frac{R^2}{\pi^2}\right)^s\zeta_R(2s-1)\nonumber\\
&&\ \ \ +\frac{2\sqrt{\pi}}{\Gamma(s)}\left(\frac{T^2}{\pi^2}\right)^s
\sum_{n=1}^\infty\sum_{p=1}^\infty\left(\frac{\pi p}{n Im \tau}
\right)^{s-1/2}K_{s-1/2}(2\pi p n Im \tau)
\ea
where $\tau=iT/R$, $\zeta_R(s)$ is the Riemann $\zeta$ function 
and $K_\nu(x)$ is a modified Bessel function. The derivative 
$\zeta_{-\de^2}^\prime(s)$ can be analytically continued to $s=0$
where it is given by
\be
\zeta_{-\de^2}^{\prime}(0)=\log(\sqrt{2R})-\frac{i\pi\tau}{12}
-\sum_{n=1}^{\infty}\log(1-q^n)
\ee
where we have defined
\be
q\equiv e^{2\pi i\tau}~.
\ee
Introducing the Dedekind $\eta$-function
\be
\eta(\tau)=q^{1/24}\Pi_{n=1}^\infty(1-q^n)\label{etadef}
\ee
we obtain finally
\be
\det(-\de^2)=\exp[-\zeta_{-\de^2}^{\prime}(0)]=
\frac{\eta(\tau)}{\sqrt{2R}}
\ee
and
\be
Z_{q}(R,T)\propto\left[\frac{\eta(\tau)}{\sqrt{R}}
\right]^{-\frac{d-2}{2}}~.\label{zetaq}
\ee
We have checked numerically that a lattice regularization of the determinant
gives the same result.
\par
Substituting in Eq.~(\ref{quantum}) we obtain \cite{ambj}
\be
<W(R,T)>=e^{-\sigma RT+p(R+T)+k}\left[\frac{\eta(\tau)}{\sqrt{R}}
\right]^{-\frac{d-2}{2}}~.
\label{prediction}
\ee
This is the prediction that we are going to compare with numerical 
results. It is important to appreciate the predictive power of the 
model: the inclusion of the fluctuation contribution does not add any 
new adjustable parameters.
\section{ The $\ZZ_2$ gauge model and the simulation algorithms}
\subsection{ The  Model}
\vskip .3 cm
The $3d$ $\ZZ_2$ gauge 
model in a cubic lattice is defined by the partition function
\eq
Z_{gauge}(\beta)=\sum_{\{\sigma_l=\pm1\}}\exp\left(-\beta S_{gauge}\right)
~.
\en
The action $S_{gauge}$ is a sum over all the plaquettes of  a cubic lattice,
\eq
S_{gauge}=-\sum_{\Box}\sigma_\Box~~~,~~~
\sigma_\Box=\sigma_{l_1}\sigma_{l_2}\sigma_{l_3}\sigma_{l_4},
\en
where $\sigma_l \in \{1,-1\}$ are Ising variables located in the links of the
lattice.

This model can be translated into the  $3d$ spin Ising 
model  by the usual Kramers-Wannier duality transformation 
\bea
Z_{gauge}(\beta)&\propto Z_{spin}(\tilde\beta)\\
\tilde{\beta}&=-\um\log\left[\tanh(\beta)\right]~~,
\eea
where $Z_{spin}$ is the partition function of the Ising model in the 
dual lattice:
\eq
Z_{spin}({\tilde\beta})=\sum_{s_i=\pm1}\exp(-\tilde\beta H_1(s))
\en
with
\eq
H_1(s)=-\sum_{\Br ij \kt}J_{\Br ij \kt}s_is_j
\en
where $i$ and $j$ denote nodes of the dual lattice and the sum is 
extended to the links ${\Br ij \kt}$ connecting  the nearest-neighbor 
sites. In general  the couplings $J_{\Br ij \kt}$ are fixed to the value $+1$ 
for all the links, but we shall use the freedom to flip some of them in the
following. 
The model is known to have a roughening transition
located at $\beta_{r}=0.47542(1)$\cite{rough_point}, 
and a deconfinement transition  at $\beta_{d}=0.7614133(22)$\cite{dec_point}.
We performed our Monte Carlo simulations  at three different values of the 
coupling constant $\beta$, all located in the rough phase and close 
enough to the deconfinement point to be well inside the scaling 
region. To avoid systematic errors due to the choice of the simulation
algorithm we used two completely different alghoritms. The first one is an
improved version of the microcanonical demon algorithm, it works directly in  
the  $\ZZ_2$ gauge  
model and we shall describe it in sect. 4.2 below. The second
one is a non-local cluster algorithm and was used to simulate
the dual spin Ising model (sect. 4.3).
The perfect agreement between the results obtained with these two 
different methods is a strong consistency check of the reliability of our data.

\subsection{The Demon Algorithm}

In our simulation the microcanonical demon algorithm of Ref.~\cite{creutz} was 
combined with a particularly efficient canonical update 
of the demons \cite{Kari} in order to obtain the canonical ensemble of the 
gauge model.
This algorithm was implemented in the multispin coding technique.
%%NEW
This means that each of the 64 bits of a word is used to store a link variable. 
The update is implemented with bit-operations. Hence 64 systems can be updated 
in parallel. Note that also the measurement of the Wilson loops is implemented
in the multispin coding technique.

The multispin coding implementation combined with the use of the 
microcanonical demon algorithm, which saves pseudo random numbers provides us 
with a performance increase compared with a naive implementation of the 
Metropolis algorithm of about a factor of 100. 

The update of a single link variable took about $2.6 \times 10^{-8} \; sec$
on a DEC250 4/266 workstation,  which is rated at 5.18 SPECint95,
6.27 SPECfp95   and 53.96 LINPACK 100 X 100 MFLOPS.

 For moderate correlation lengths, as discussed in this paper,
 the microcanonical demon algorithm
 should  provide superior performances compared to a 
 cluster-update \cite{bkkls}. 

%%NEW
 For a detailed discussion see Ref.~\cite{glue} where the 
 same algorithm was used.
%%ENDNEW

\subsection{Non-local Cluster Algorithm}

 Using the duality transformation it is possible to build 
up a one-to-one mapping of physical observables of the gauge system into 
the corresponding spin quantities. For instance, the vacuum 
expectation value of a Wilson loop $W(C)$ can be expressed in terms of 
spin variables as follows. First, choose a surface $\Sigma$ bounded 
by $C$: $\partial\Sigma=C$; then  take $J_{\Br ij \kt}=-1$ 
for all those links intersecting $\Sigma$ and denote with $H_{-1}(s)$ the 
Ising Hamiltonian with this choice of couplings. The new Ising partition 
function 
\be
Zw_{spin}({\tilde\beta})=\sum_{s_i=\pm1}\exp\left(-\tilde\beta 
H_{-1}(s)\right)
\ee
describes a vacuum modified by the  Wilson loop $W(C)$; we call it the 
W-vacuum (WV). It is easy to see that  
\eq
\Br W(\partial \Sigma)\kt_{gauge}={Zw_{spin}\over Z_{spin}}=
\Br\prod_{\Br ij\kt\in\Sigma}
\exp(-2{\tilde\beta}s_is_j)\kt_{spin}~~~,
\label{wil}
\en
where  the product is over all the dual links intersecting $\Sigma$.

For large Wilson loops there is some problem in calculating the 
above expectation value, because each term inside the brackets is the 
product of factors which can be very large or very small, then one 
expects large fluctuations. There is however a new method where this 
difficulty is overcome by adapting a new procedure first applied by 
one of us \cite{mh} to the calculation of the surface tension. 
The idea is to 
consider the sign $w=\pm1$ of the couplings settling the presence or the 
absence of the Wilson loop in the vacuum as a dynamical variable and sum 
over it. Thus the new partition function is now
\eq
Z=\sum_{w=\pm1}\sum_{s_i=\pm1}\exp\left(-\tilde\beta H_w(s)
\right)~~~.
\en
The fraction of configurations with the Wilson loop is given by the 
ratio
\eq
\frac{Zw}{Z}={\sum_{w=\pm1}\sum_{\strut{s_i=\pm1}}
\exp(-\tilde\beta H_w(s))\delta_{w,-1}\over Z}=\Br\delta_{w,-1}\kt
\en
while the fraction of configurations corresponding to the ordinary 
vacuum is $\Br\delta_{w,1}\kt$. The comparison with Eq.~(\ref{wil})
yields
\eq
\Br W(\partial \Sigma)\kt_{gauge}=\frac{\Br\delta_{w,-1}\kt}
{\Br\delta_{w,1}\kt}~~~,
\en
hence the evaluation of this vacuum expectation value is brought back 
to mediate over quantities which take only 0 or 1 values. 

A great advantage of mapping a gauge observable into one of a spin 
system is that it can be used,  in place of the 
usual Metropolis or heat-bath methods, a  non local cluster 
updating algorithm \cite{sw}, which has been proven very successful in 
fighting critical slowing down. 
For instance, the Swendsen-Wang cluster algorithm \cite{sw} is made
of two steps:

\begin{description}
\item{$\bullet$} Some bonds  are deleted 
with the following rule: 
the bond ${\Br ij \kt}$ is deleted
with probability 
\eq
p=\cases{\exp(-2\tilde\beta), & if $s_i=s_j$\cr
1~~~,& if $s_i\not=s_j$\cr}
\en

In this way the lattice is split into clusters of sites connected by 
the  remaining (or frozen) bonds; these form the so called Kasteleyn-
Fortuin (KF) clusters \cite{kf}.
\item{$\bullet$}
These clusters are randomly flipped with probability one half.
\end{description}
This updating algorithm provides us with a very efficient method to 
estimate $\Br W(\partial \Sigma)\kt_{gauge}$. 
It is sufficient to look at the KF 
clusters which intersect $\Sigma$ in each MC configuration: if at 
least one of these clusters is topologically linked to the loop 
$C=\partial \Sigma$ with an odd winding number, then the MC configuration is 
incompatible with the presence of the Wilson loop (hence $\delta_{w,-1}=0$). 

On the contrary, if all these clusters are not linked to $C$, the 
configuration is compatible with the presence of $W(C)$: the clusters 
intersected by $\Sigma$ are split into an upper and a lower part. We can  
flip all the spins of the lower parts and the signs of the couplings
$J_{\Br ij\kt}$ cut by $\Sigma$, transforming a configuration of the 
ordinary vacuum into one of the W-vacuum. 
Conversely, if the MC configuration is one of the W-vacuum,  
there is no cluster linked to $C$ and the configuration can be always 
transformed, within the same flipping procedure, into one of the ordinary 
vacuum. Because of such a property of the WV configurations, it is not
actually necessary to  flip the $J_{\Br ij \kt}$ couplings, and the 
evaluation of $\Br W(C)\kt$ is reduced to study the winding number
of the KF clusters around the loop $C$ \cite{gv} : denoting with $N_W$ 
the number of MC configurations compatible with  a W-vacuum and with $N$ 
the total number of iterations we have simply
\eq
\Br W(\partial\Sigma)\kt=\frac{N_W}{N}~~~.
\en
\section{Comparison with Monte Carlo data}
In this section we compare the prediction of the effective string theory with
Monte Carlo data for the $3d$ $\ZZ_2$ gauge model. Including the 
effective
string contributions computed in Sec. 2 the functional dependence 
of the Wilson loop on the sides $R,T$ is, for $d=3$, 
\be
\Br W(R,T)\kt =\exp\left[-\sigma T R+p(T+R)+k\right]
\left[\frac{\eta(\tau)}{\sqrt{R}}\right]^{-1/2}~.
\ee
The values of the parameters $\sigma$, $p$ and $k$ are not predicted by 
the effective string theory. However, we can eliminate $p$ and $k$ by
considering ratios of Wilson loops with equal perimeter.
Moreover, the value of $\sigma$ can be taken from high precision numerical 
simulations of the $3d$ Ising spin model, which is dual to the 
$\ZZ_2$ gauge
model. The string tension of the gauge model corresponds, by duality, to the
interface tension of the spin model. Using these values for $\sigma$ we achieve
complete independence between the theoretical predictions and the numerical data
to be compared.\par
In particular, defining
\be
R(L,n)\equiv\frac{\Br W(L+n,L-n)\kt}{\Br W(L,L)\kt}\exp(-n^2\sigma)
\ee
the string theory prediction for $R(L,n)$ depends only on $t=n/L$:
\be
R(L,n)=F(t)=\left[\frac{\eta(i)\sqrt{1-t}}{\eta\left(i\frac{1+t}{1-t}\right)}
\right]^{1/2}\label{ft}
\ee
and does not contain any adjustable parameters.

For  $\beta=0.75245$ we 
used the non--local cluster algorithm 
(Subsection 4.3). Non--local algorithms drastically reduce correlations
in Monte Carlo time, therefore we could
safely perform a measurement for every configuration we produced. 
On the other hand, the nature of the algorithms is such that only one 
ratio of Wilson loops can be measured in each run. For each ratio about
$5\cdot 10^6$ iterations were performed.\par
For $\beta=0.75202$ and $\beta=0.75632$ we used the demon algorithm 
described in Subsection 4.2.
The run at $\beta=0.75632$ on the $80^3$ lattice (from which most
of the data reported in this paper were taken) took 6.5 days on
a Alpha AXP-3000/400. We performed a measurement after 50 updating sweeps.
The integrated autocorrelation time (in units of sweeps)
was smaller then 300 for all  Wilson loops considered.

For technical reasons in the simulations at $\beta=0.75202$ and
$\beta=75245$ we chose asymmetric
lattices with different sizes $N_s$ and $N_t$ in the space and time directions
respectively. For these value of $\beta$ we only measured Wilson loops
 of  sizes $R\times T$ 
orthogonal to the time direction.  $R$ and $T$ were
constrained to be: $\xi<<R,T<<N_s$ where $\xi$ is the bulk correlation length.

A summary of the informations on our data sample is reported in tab.I~.

 \begin{table}[ht]
 \caption{\sl  Some informations on the data sample:
$\beta$ is the coupling constant of the gauge $\ZZ_2$
 model and $\tilde\beta$ the
corresponding (dual) coupling for the spin model. ``MD'' denotes the
microcanonical demon algorithm, while 
``NL'' denotes the nonlocal one. $N_t$ and $N_s$
 are the lattice
sizes. 
With $N_c$ we denote the total amount of configurations in the run. $\xi$
denotes the bulk correlation length (taken from ref~[12] and [16].}
 \label{data1}
  \begin{center}
   \begin{tabular}{|c|c|c|c|c|c|}
   \hline
 $\beta$ & $\tilde\beta$ & {\bf alg.}& $N_t\times N_s^2$ & $N_c$ & $\xi$ \\
   \hline
 0.75202 &  0.22600& MD & $ 70\times 50^2$ & $2\cdot 10^6$  &  3.135(9)   \\
 0.75245 &  0.22580& NL & $ 24\times 30^2$ & $5\cdot 10^6$ &  3.170    \\
 0.75632 &  0.22400& MD & $ 80^3$ & $1.7\cdot 10^5$ & 4.64(3)    \\
   \hline
   \end{tabular}
  \end{center}
 \end{table}

\par
For each value of 
the asymmetry ratio $t$, we 
have displayed in Tab. \ref{tab:compari} the value of 
$R(L,n)$ corresponding to the largest value of the physical size 
$\sigma L^{2}$ available in our data sample. 
The values of the string tension are taken from Ref.~\cite{cfghpv} ($\beta=
0.75245$ and $\beta=0.75632$) and \cite{hp} ($\beta=0.75202$).
The comparison with the
prediction of the effective string theory gives a reduced $\chi^2$ of
1.2. 
The same data are plotted 
in Fig. 1.

\begin{table*}[hbt]
% space before first and after last column: 1.5pc
% space between columns: 3.0pc (twice the above)
\setlength{\tabcolsep}{0.97pc}
% -----------------------------------------------------
% adapted from TeX book, p. 241
\newlength{\digitwidth} \settowidth{\digitwidth}{\rm 0}
\catcode`?=\active \def?{\kern\digitwidth}
% -----------------------------------------------------
\caption{\sl Monte Carlo results for the ratios $R(L,n)$}
\label{tab:compari}
\begin{tabular*}{\textwidth}{lrlllr}
\hline
$n/L$&$L$&$\ \ \beta$&$\ \ \sigma$&$R(L,n)$&$F(n/L)$\\
\hline
$1/5\ $   &$15$&$0.75202$&$0.01023(5)$& $1.0104(17)$&$1.00881$\\
$1/4\ $  &$20$&$0.75632$&$0.004779(14)$&$1.0166(10)$&$1.01453$\\
$2/7\ $ &$ 7$&$0.75245$&$0.009418(61)$&$1.0229(24)$&$1.01987$\\
$1/3\ $  &$12$&$0.75632$&$0.004779(14)$&$1.02881(30)$&$1.02901$\\
$3/8\ $ &$ 8$&$0.75245$&$0.009418(61)$&$1.0403(31)$&$1.03940$\\
$9/20\ $  &$20$&$0.75632$&$0.004779(14)$&$1.0684(23)$&$1.06588$\\
$1/2\ $   &$20$&$0.75632$&$0.004779(14)$&$1.0911(27)$&$1.09153$\\
$3/5\ $   &$25$&$0.75632$&$0.004779(14)$&$1.165(11) $&$1.17667$\\
\hline
\end{tabular*}
\end{table*}
\begin{figure}
\begin{center}
\mbox{\epsfig{file=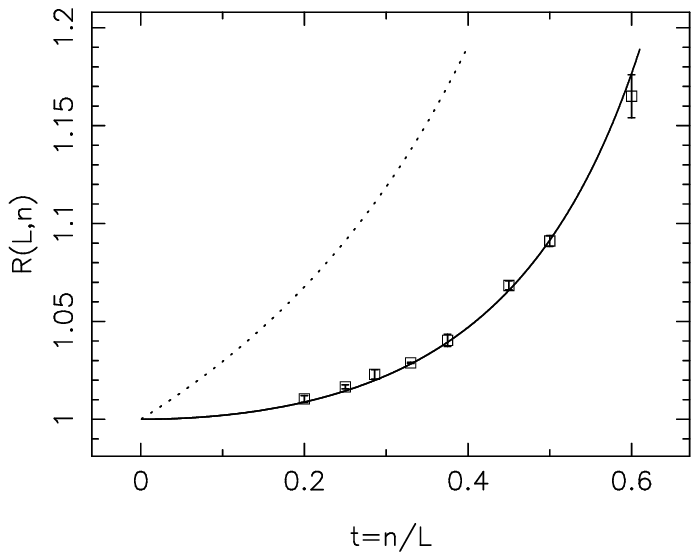}}
\vskip 2mm
\caption{Comparison of the prediction of the free string model with Monte Carlo
data for the ratios $R(L,n)$. The solid line is the prediction of the 
string model, Eq.~(\ref{ft}). The dotted line is the same prediction when
only the conformal anomaly term is taken into account, Eq.~(\ref{f0}).}
\label{fig:compa}
\end{center}
\end{figure}
Two facts emerge immediately from the data:
\begin{description}
\item{a)}The contribution of the flux--tube fluctuations are 
{\em quantitatively} relevant in the physics of confinement: notice 
that a simple perimeter-area law would predict all ratios $R(L,n)$ to 
be equal to one.
\item{b)} For large enough Wilson loops, the effective bosonic string 
model, Eq.~(\ref{z1loop}), describes these fluctuations with great 
accuracy.
\end{description}
It is important to notice that the whole functional form of the 
string contribution to the Wilson loop must be taken into account to 
obtain a satisfactory agreement with the numerical data. In particular 
it is a rather common practice to take into account only the conformal 
anomaly term, {\em i.e.} to write 
\be
Z_q(R,T)\propto q^{-\frac{d-2}{48}}\label{confonly}
\ee
corresponding in our case to 
\be
R(L,n)=F_0(t)=\exp\left(\frac{\pi}{12}\frac{t}{1-t}\right)~.\label{f0}
\ee
It is apparent from Fig.\ref{fig:compa} that this approximation (dotted 
line) is not adequate, and can result in serious errors in extracting the
string tension from a set of Wilson loop expectation values. The largest 
part of the discrepancy is due to the $\sqrt{R}$ term in Eq.~(\ref{zetaq})
rather than to the subdominant terms in the expansion of $\eta(\tau)$
in powers of $q$.
\vskip0.5cm
If we consider Wilson loops smaller than a threshold size of order 
$\sigma L^{2}\sim 1$, rather significant finite--size effects appear. 
An example of these is shown in Fig. 2, where we have plotted the 
values of $R(L,n)$ for a fixed asymmetry ratio $t=1/2$ and 
different values of the physical size $\sigma L^{2}$ of the Wilson 
loop. The fact that these effects appear to depend on the physical 
size $\sigma L^{2}$ rather than on the size expressed in lattice 
units suggests that they are related to physical effects rather than 
lattice artifacts.
An important lesson can be drawn from these data: the free string 
model is indeed valid in the infrared limit only. If a data sample 
consists of too small Wilson loops, one can erroneously conclude that 
the model is not accurate.
\begin{figure}
\begin{center}
\mbox{\epsfig{file=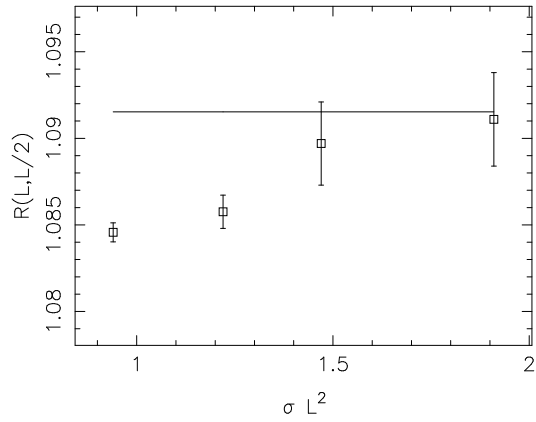}}
\vskip 2mm
\caption{Finite--size effects for small Wilson loop.
The prediction of the free string model is $R(L,L/2)=1.09153\dots$
(straight line).}
\label{fig:finite}
\end{center}
\end{figure} 
\section{Conclusions}
Our present knowledge about the effective string picture of confinement in 
LGT can be summarized in the following statements:
\begin{itemize}
\item{} The idea of a massless, fluctuating flux--tube joining 
static sources in the rough phase of LGTs is not only an intuitive 
physical picture, but has precise and very predictive quantitative 
consequences, which can be tested with numerical simulations.
\item{} The contribution of the flux--tube fluctuations to the 
Wilson loop are numerically relevant: failure to take them into 
account would result in a biased analysis of the data.
\item{} In the infrared regime, for the $3d$ $Z_{2}$ gauge model, we 
have shown that the fluctuations are accurately described by an 
effective bosonic string theory. The high degree of precision 
attainable in this model allows us this precise identification of the 
effective string theory. 
\item{} However, since it is widely believed that the infrared 
behavior of LGTs is largely independent of the particular gauge model 
considered, we conjecture that the same effective string 
theory will describe the flux--tube fluctuations in a large class of 
confining LGTs.
\item{} This effective theory is truly an infrared limit: small 
Wilson loops show rather large finite--size effects. The effective 
model can be applied only when data for sufficiently large (in 
physical units) Wilson loops are available.
\end{itemize}
\vskip0.5cm
It is the availability of {\em precise} data on {\em large} Wilson 
loops that makes the precise identification of the effective string 
theory possible. For example in Ref.~\cite{fermionic} a different 
effective string theory, of fermionic nature, 
was shown to describe the flux--tube 
fluctuations as accurately (and in some cases even more accurately) 
than the bosonic theory. The dramatically increased precision on large 
Wilson loops allows us now to identify the bosonic model as the most 
accurate.

\vskip .3 cm
We would like to thank K. Pinn and S. Vinti for many useful discussions.
One of us (F.G.) would like to thank Juan A. Mignaco for fruitful discussions 
and the Instituto de F\'isica of the Universitade Federal do Rio de Janeiro 
for the warm hospitality during the completion of this work.
This work has been supported in part by the European Commission TMR 
programme ERBFMRX-CT96-0045 and by the Ministero ita\-lia\-no 
del\-l'Uni\-ver\-si\-t\`a e della Ricerca Scientifica e Tecnologica.


\begin{thebibliography}{99}

\bibitem{conj}H.B. Nielsen and P. Olesen, \NP{B61} (1973) 45. 
G.'t Hooft, \NP{B72} (1974) 461.
\bibitem{wilson} K. Wilson \PR{D10} (1974) 2445.
\bibitem{polya} A.Polyakov, \PL{82B} (1979) 247; \NP{B164} (1980) 171
\bibitem{gene} J-L. Gervais and A. Neveu, \PL{80B} (1979) 255.
\bibitem{nambu} Y. Nambu, \PL{80B} (1979) 372.
\bibitem{mm} Y.Makeenko and A. Migdal, \NP{B188} (1981) 269;
I. Kostov, \NP{B265} (1986) 223.
\bibitem{awa} M. Awada and D. Zoller, \PL{B325} (1994) 115 
(hep-th/9404077);
D.V. Antonov and D. Ebert, preprint hep-th/9608072.
\bibitem{pol} A.M. Polyakov, preprint hep-th/9607049.
\bibitem{gross} D. Gross and W. Taylor, \NP{B400} (1993) 181 
(hep-th/9301068); 
\NP{B403} (1993) 395 (hep-th/9303046).
\bibitem{rough} A. Hasenfratz, E.Hasenfratz and P. Hasenfratz \NP{B180} (1981)
353.
C. Itzykson, M.E.Peskin and J.B.Zuber, \PL{95B} (1980) 259.
\bibitem{lsw} M. L\"uscher, K. Symanzik and P. Weisz, Nucl. Phys. 
{\bf B173} (1980) 365.
M. L\"uscher, Nucl. Phys. {\bf B180} [FS2] (1981) 317.
\bibitem{cfghpv} M. Caselle, R. Fiore, F. Gliozzi M. Hasenbusch, 
K. Pinn and S. Vinti, \NP{B432} [FS] (1994) 590 (hep-lat/9407002).
\bibitem{lmw} M. L\"uscher, G. M\"unster and P. Weisz,\NP{ B180}
[FS2] (1980) 1.
\bibitem{width} M. Caselle, F. Gliozzi, U. Magnea and S. Vinti,
\NP{B460} (1996) 397.
\bibitem{cft} C. Itzykson and J.-M. Drouffe, ``Statistical Field 
Theory'', Cambridge 1989, Ch. 9.
\bibitem{glue} V. Agostini, M. Carlino,  M. Caselle and M. Hasenbusch,
preprint hep-lat/9607029, to be published on Nuc. Phys. B. 
\bibitem{pv2}P. Provero and S. Vinti, Physica {\bf A 211} (1994) 436
(hep-lat/9310028).
\bibitem{pv1}P. Provero and S. Vinti, Nucl. Phys. {\bf B 441} (1995) 
562 (hep-th/9501104).
\bibitem{ambj} J. Ambjorn, P. Olesen and C. Peterson, \NP {\bf B 
244} (1984) 262.
\bibitem{olesen}P. Olesen, Phys. Lett. {\bf B 160} (1985) 144.
\bibitem{creutz} M. Creutz,  Phys. Rev. Lett.  {\bf 50} (1983) 1411, \\
     M. Creutz, G. Bhanot and H. Neuberger,
      Nucl. Phys. {\bf B 235 [FS11]} (1984) 417
\bibitem{Kari} K. Rummukainen,  Nucl. Phys. {\bf B 390} (1993) 
621 (hep-lat/9209024).
\bibitem{bkkls} R. Ben-Av, D. Kandel, E. Katznelson, P.G. Lauwers and S.
Solomon,  J. Stat.Phys. {\bf 58} (1990) 125
\bibitem{mh} M.Hasenbusch,J.Phys {\bf I 3} (1993) 753 ; 
Physica {\bf A 197} (1993) 423.
\bibitem{sw} R.H.Swendsen and J.S. Wang,\PRL{58} (1987) 86
\bibitem{kf}P.W.Kasteleyn and C.M.Fortuin, J. Phys. Soc. Japan (suppl) 
{\bf 26} (1969) 11.
\bibitem{rough_point} M. Hasenbusch and K. Pinn, preprint HUB-EP-96/12,
MS-TPI-96-8, cond-mat/9605019.
\bibitem{dec_point} M.W.J. Bl\"ote, E. Luijten and J.R. Heringa, 
J. Phys. A: Math. Gen. 28 (1995) 6289.
\bibitem{hp} M. Hasenbusch and K. Pinn, Physica {\bf A 192} (1993) 342.
\bibitem{gv} F.Gliozzi and S. Vinti, contribution to Lattice'96
\bibitem{fermionic}M. Caselle, R. Fiore, F. Gliozzi, P. Provero and S. 
Vinti, \IJMP{\bf A 6}(1991) 4885.\\ M. Caselle, R. Fiore, F. Gliozzi,
P. Guaita and S. Vinti, \NP {\bf B 422} (1994) 397 (hep-lat/9312056).
\end{thebibliography}
\end{document}